\documentclass[12pt,preprint]{aastex}

\begin{document}

\title{The large-scale $rms$ bulk velocity estimated from
   QSOs' Ly$\alpha$ forests}

\author{Hu Zhan and Li-Zhi Fang}

\affil{Department of Physics, University of Arizona, Tucson,
AZ 85721}

\begin{abstract}

We propose a method for estimating the large-scale $rms$ bulk velocity
of the cosmic mass field from the transmitted fluxes of Ly$\alpha$ forests.
It is based on two linear relationships on large scales: 1) the relation
between the fluctuations of the transmission and the underlying density
field, and 2) the relation between the density fluctuations and the 
peculiar velocity field. We show that, with a multiscale decomposition, 
the two relations can be effectively employed for predicting the $rms$ bulk
velocity. Since QSO's Ly$\alpha$ forest is due to the absorptions of diffusely
distributed and photoionized IGM, this method provides an independent
estimate of the $rms$ bulk velocity at high redshifts, on large scales,
and free from the bias of galaxies. Using the transmitted flux of 60 
moderate-resolution QSO spectra, the $rms$ bulk velocity is found to be 
230$\pm$50 km s$^{-1}$ around redshift $z=2.25$ on scale 23 h$^{-1}$Mpc, 
and down to 110$\pm$45 km s$^{-1}$ around $z=3.25$ on scale 92 h$^{-1}$Mpc
for an LCDM universe ($\Omega=0.3$ and $\Lambda=0.7$). The results are 
basically consistent with the linear evolution theory.

\end{abstract}


\section{Introduction}

The transmitted fluxes of high redshift QSOs' Ly$\alpha$ absorption
spectra have been successfully applied to many aspects of cosmic
large-scale structure study, such as the discrimination among dark
matter models (Bi, Ge \& Fang 1995; Bi \& Davidsen 1997), recovery of
the initial linear mass power spectrum and estimation of cosmological
parameters (Croft et al. 1998 \& 1999; Hui 1999; Nusser \& Haehnelt
1999; McDonald \& Miralda-Escud\'e 1999; Feng \& Fang 2000; McDonald
et al 2000), and finding the applicable range of the hierarchical
clustering model (Feng, Pando \& Fang 2001; Zhan, Jamkhedkar \& Fang
2001). These studies show that Ly$\alpha$ forests provide an important
complement to structure formation studies based on galaxy samples.

In this paper we extend the application of Ly$\alpha$ forests to the
cosmic velocity field. In terms of cosmological parameter determination,
the $rms$ bulk velocity is an important statistic, as it is related to an
integrated power spectrum over the scales beyond the directly detectable
range of the power spectrum with current galaxy surveys.
On such scales, the cosmic clustering remains in the linear regime,
and, therefore, would be useful to set a constraint on cosmological
parameters, and on the relation between the galaxy distribution and the
underlying mass field.

Many efforts have been made on measuring and estimating cosmic bulk
velocity with galaxy samples. An important result is that the velocity
field calculated from spatial distribution of galaxies is found to be
basically consistent with the peculiar velocities measured based on
the Tully-Fisher or fundamental plane relations (e.g. Dekel et al
1999; Branchini et al 1999).  This indicates that the bulk velocity of
galaxies on large scale is of gravitational origin. That is, the
cosmic velocity field ${\bf v}({\bf x})$ on large scales is related to
the cosmic mass density field $\delta({\bf x})$ via the following
linear relation or its variants
\begin{equation}
\delta({\bf x}) = - \frac{1}{H_0f} \nabla \cdot {\bf v}({\bf x}),
\end{equation}
where $f\simeq \Omega^{0.6}$ for the local universe. We define
${\bf x}=(x,y,r)$, where $x$ and $y$ are the coordinates on the celestial
sphere, and $r$ is the radial or redshift direction.

A problem with galaxy samples, however, arises from the likely bias of
the distributions of galaxies relative to the underlying mass
distribution in the universe.  The simplest bias model assumes
that the number density fluctuations of galaxies, $\delta_g({\bf x})$,
is related to the underlying mass density field by $\delta_g({\bf
x})=b\delta({\bf x})$, where $b$ is the bias parameter. More
sophisticated models assume that the relation between $\delta_g({\bf
x})$ and $\delta({\bf x})$ might be stochastic, non-local, and
non-linear. Furthermore, the peculiar velocities of galaxies may also
be biased from the cosmic velocity field.  These effects will cause
uncertainty in deriving the cosmic velocity field from galaxy
samples. It is therefore important to develop a method of estimating
the cosmic velocity field free from galaxy bias.

Here we calculate, based on Eq. (1), the large-scale $rms$ bulk
velocities from samples of the transmitted flux $F$ of QSO's
Ly$\alpha$ absorption spectrum. The QSO Ly$\alpha$ forest is very well
modeled by absorption from diffusely distributed and photoionized IGM,
which in turn is believed to be distributed proportionally to the
density of the underlying mass field $\rho({\bf x})$ on scales larger
than the Jeans length of the IGM (for a review, see Rauch 1998). The
transmitted flux of QSO's Ly$\alpha$ forest is thus likely to be less
biased as a tracer of the underlying mass and velocity field,
providing an independent estimate of the $rms$ bulk velocity,
separate from that defined by galaxies.

To do so, we need to develop methods for: 1) Obtaining mass
fluctuation data $\delta({\bf x})$ from Ly$\alpha$ flux $F$ in QSO
spectra, in the linear or quasi-linear regimes.  2) Properly
calculating the bulk velocity from mass fluctuations.  These two methods
are discussed, respectively, in \S 2 and \S 3. The results of our
investigation of the $rms$ bulk velocity on large scales at high
redshifts are presented in \S 4.

\section{Probability Distribution Function of Ly$\alpha$ Transmission}

\subsection{Transmission--density relation}

For a QSO with coordinate $(x_i, y_i)$ on the celestial sphere,
the observed flux of Ly$\alpha$ absorptions is equal to $F_ce^{-\tau(z)}$,
with $F_c$ being the continuum, and $\tau(z)$ the optical depth. The 
data to be used have been normalized, and therefore the transmission 
$F(z)=e^{-\tau(z)}$.

For a diffusely distributed HI cloud in photoionization equilibrium,
we have (e.g. Bi 1993; Fang et al. 1993)
\begin{equation}
\tau(z)=A(z)\int dr [1+\delta(x_i,y_i,r)]^a V[w(z)-r-v_r(x_i,y_i,r), b]
\end{equation}
where $w(z)$ is the redshift space coordinate, $r$ is the radial
comoving coordinate, $V$ is the normalized Voigt profile that is
approximately Gaussian, $V \simeq 1/(\sqrt{\pi} b) \times
\exp\{-[w(z)-r-v_r(x_i,y_i, r)]^2/b^2\}$, $v_r(x_i,y_i,r)$ is the
radial component of peculiar velocity, $b$ is the Doppler broadening
on the order of several tens km s$^{-1}$, and
$\delta(x_i,y_i,r)=[\rho(x_i,y_i,z)-\bar{\rho}]/\bar{\rho}$ is the
dark matter density contrast smoothed on the scale of the IGM Jeans
length (e.g. Bi \& Davidsen 1997; Nusser \& Haehnelt 1999).  The
parameter $A(z)$, which depends on the cosmic baryonic density, the
photoionization rate of HI, and the mean temperature of IGM, is
referred to the mean transmission over a redshift range as $\langle
F(z) \rangle =\langle e^{-\tau(z)} \rangle = e^{-A(z)}$. The parameter
$a$ depends on reionization, and is in the range of 1.5 to 1.9 (Hui \&
Gnedin 1997).

There are two problems in drawing information of linear or quasi-linear
density fluctuations $\delta$ from the transmitted flux $F$. First, the
relation between $F$ and $\delta$ is non-linear. This is especially
true for a saturated absorption line, i.e. $F \simeq 0$ corresponds
to the non-linear regime of $\delta$. Second, the peculiar velocity
$v_r$, which is what we are trying to detect, already enters in Eq.(2).
These two problems can be avoided by considering the smoothed $F$
and $\delta$ on large scales.  If the smoothing scales are much larger
than the Doppler factor $b$ and the positional uncertainty caused by
the peculiar velocity $v(x)$, the Voigt function $V$ can be
approximated as a Dirac delta function $\delta^D(r-w(z))$.
In this case, Eq.(2) yields
\begin{equation}
F(z) \simeq \exp \{ - A(z)[1+\delta(x_i,y_i,z)]^a \}.
\end{equation}
For simplicity, in Eq.(3) and hereafter we still use the notations
$F(z)$ and $\delta(z)$ for the smoothed transmission and density
perturbations. Thus, for the linear or quasi-linear regime, i.e.,
$\vert \delta F(z) \vert = \vert F(z)-\langle F(z)\rangle \vert /
\langle F(z)\rangle < 1$ and $\vert \delta(x_i,y_i,z)\vert < 1$,
Eq.(3) gives
\begin{equation}
\delta(x_i,y_i, z)  \simeq  -\frac{1}{aA(z)}\delta F(z),
\end{equation}
where $(x_i,y_i)$ is to remind that it is for the $i$-th QSO.

Eq.(4) shows that the fluctuations of the Ly$\alpha$ transmission flux
$\delta F(z)$ trace the underlying density perturbation
$\delta(x_i,y_i,z)$ point-by-point. The factor $aA(z)$ plays the role
of a bias parameter between the Ly$\alpha$ flux fluctuations and the
mass density perturbations. $aA(z)$ is of order of 1, as $a\simeq
1.5-1.9$ and $A(z)\simeq 0.75-0.9$. Eq.(3) has been used in the
recovery of the initial linear power spectrum of the underlying mass
field by the observed Ly$\alpha$ transmission (e.g. Croft et
al. 1998). It should be reminded that Eq.(4) holds only if 1) the
variance of $\delta F$ is less than 1, or the variance of $F$ is less
than $\langle F(z)\rangle$; and 2) the positional uncertainty caused
by the peculiar velocity is much smaller than the scales
considered. We will check the validity of these conditions in the
analysis below.

\subsection{Transmission--density relation in DWT representation}

To do the smoothing of the transmitted fluxes, we apply the algorithm of
multiscale-decomposition based on the discrete wavelet transform (DWT)
(Daubechies 1992; Fang \& Thews 1998). Because the bases of the DWT
analysis are orthogonal, complete and localized, it is easy for the
calculation to use Eq.(1) (next section). In our calculation, we use
the Daubechies 4 (D4) wavelets.

Let us consider a sample $\delta F(z)$ that spans a redshift space
$L=z_2-z_1$, and the spectrum is binned into $2^J$ pixels with $J$ being an
integer. With the DWT analysis, the smoothed $\delta F(z)$ on the scale
$L/2^j$ is given by
\begin{equation}
\delta F_j(z)=\sum_{l=0}^{2^j-1} \epsilon^F_{jl}\phi_{jl}(z),
\end{equation}
where $\phi_{jl}(z)$s are the DWT scaling functions for mode $(jl)$.
The label $j$ is for spatial scale $L/2^j$, and $l$ is for the
position in redshift space around $z_1+lL/2^j$. The scaling function
plays the role of a window function. The scaling function coefficient
(SFC) $\epsilon^F_{jl}$ is given by
\begin{equation}
\epsilon^F_{jl}(x_i,y_i)=\int \delta F(z)\phi_{jl}(z)dz,
\end{equation}
where $(x_i,y_i)$ is to show that the flux fluctuation $\delta F(z)$ 
in Eq.(6) is for the $i$-th QSO. The SFC $\epsilon^F_{jl}$ is 
proportional to the mean flux over a bin size $L/2^j$ around the 
position $l$. Thus, with the DWT decomposition, Eq.(4) becomes
\begin{equation}
\epsilon^{m}_{jl}(x_i,y_i) = -\frac{1}{aA(z)}\epsilon^F_{jl}(x_i,y_i),
\end{equation}
where the SFC of the mass field is given by
\begin{equation}
\epsilon^{m}_{jl}(x_i,y_i)=\int \delta(x_i,y_i, z)\phi_{jl}(z)dz.
\end{equation}
Therefore, Eq.(7) is the DWT representation of Eq.(4).
The SFCs $\epsilon^F_{jl}$ and $\epsilon^{m}_{jl}$ actually are,
respectively, the fluctuations $\delta F(z)$ and $\delta(z)$ in
DWT representation. Eq.(7) has been used in the DWT recovery of
the initial linear power spectrum of the underlying mass field by
the observed Ly$\alpha$ transmission flux (Feng \& Fang 2000).

\subsection{PDF of Ly$\alpha$ Transmission}

The samples of transmitted fluxes $F(z)$ used in our study comprise
of 60 QSOs' spectra selected from Bechtold (1994),
Dobrzycki \& Bechtold (1996), Scott, Bechtold \& Dobrzycki (2000),
and Scott et al. (2000). The selection ensures that each spectrum 
in the calculation below is comparable to at least twice the length 
of the largest scale (92 h$^{-1}$Mpc) interested.
The emission redshift of the QSOs covers the
range from 1.9 to 4.12. Each spectrum is averaged in bins of size
$\sim 1.2 \AA$, such that there are 1024 `pixels' per unit redshift,
which converts to a resolution of $c/1024 \simeq 293$ km s$^{-1}$.

Using Eq.(6), we compute the SFCs $\epsilon^F_{jl}$ of flux fluctuations
$\delta F(z)$ for each QSO. We use the so-called off-counting method
to treat the spectra (Jamkhedkar, Bi \& Fang 2001), i.e. the SFC
$\epsilon^F_{jl}$ of modes $(j,l)$ is not counted in the calculation
if the data at that mode are contaminated by metal line, bad, or low S/N
pixels. An advantage of the DWT off-counting method is that the
statistical results on large scales are not sensitive to the bad pieces
on small scales.

We treat all the data as ensembles of various modes $(j,l)$. That is,
for a given mode $(j,l)$, the ensemble consists of all SFCs
$\epsilon^F_{jl}(x_i,y_i)$ of the QSOs that have flux observed around
$\lambda = (1+z_1+lL/2^j)\lambda_{\alpha}$, where
$\lambda_{\alpha}\simeq 1216\AA$. Thus, we can calculate the ensemble
average of SFCs of each mode $(j,l)$ by
\begin{equation}
\epsilon^F_{jl}=\frac{1}{N}\sum_{i=1}^{N} \epsilon^F_{jl}(x_i,y_i).
\end{equation}
where $N$ is the number of the SFCs in the ensemble.

Since modes $(j,l)$ are localized in redshift space, we can also
construct ensembles of the SFCs in different redshift ranges. For
a given redshift range $z \pm (\Delta z/2)$, the ensemble consists
of all $\epsilon^F_{jl}(x_i,y_i)$ of which the label $l$
corresponds to position in the range $z \pm (\Delta z/2)$. In the
following calculations, we use three redshift ensembles. They are
$z$=2.25, 2.75, and 3.25 with $\Delta z=0.50$.

The probability distribution functions (PDFs) of the SFCs
$\epsilon^F_{jl}$ are calculated in the three redshift ranges. The
result for $z = 2.75$ is plotted in Figure 1. The scale $j$ corresponds
to redshift distance $\delta z= \Delta z /2^{j}=2^{-j-1}$. In Fig. 1,
the physical scale of $j$ is calculated from the redshift distance
$\delta z$ at $z =2.75$, for a LCDM model ($\Omega = 0.3$ and
$\Lambda = 0.7$). For instance, the scale 92 h$^{-1}$ Mpc corresponds
to $\delta z = 0.125$, or in wavelength $\Delta \lambda \simeq 152 \AA$.

In Figure 1, the panel on the smallest scale, i.e. 72 h$^{-1}$kpc, is
the PDF of the binned distribution $F(z)$ without the SFC smoothing,
and therefore, it actually is the PDF of QSO Ly$\alpha$ transmission,
which has been done by many others (e.g. McDonald et al. 2000).
Rather, Figure 1 shows the scale dependence of the
PDFs.  When the smoothing scale is small (large $j$), the PDF is
significantly non-Gaussian, while it is essentially Gaussian on large
scales. The K-S test shows that the PDFs are Gaussian on scales larger
than 12 h$^{-1}$ Mpc, and non-Gaussian on scales less than
$\sim 12$ h$^{-1}$ Mpc. This is consistent with the
authors' previous work (Zhan, Jamkhedkar, \& Fang 2001). Moreover,
the variance of $2^{(j-9)/2}\epsilon^F_{jl}$, or equivalently the
variance of $\delta F_j(z)$, is less than 1. Thus, it is reasonable
to calculate the density perturbations of the underlying mass field
by Eqs.(4) or (7).

All the above-mentioned properties are typical. That is, in other
redshift ranges, the behavior of the $\epsilon^F_{jl}$ PDFs is similar
to Figure 1 with a slight evolution in the redshift range considered.
Thus, on scales larger than 12 h$^{-1}$ Mpc, one can calculate
the SFCs of the mass field with Eq.(7). In these cases, Eq.(7) also
insures that the variance of the density fluctuations $\delta$ is less
than 1, and therefore, the corresponding evolution of the mass field
is still in the linear or quasi-linear regime.

\section{Ly$\alpha$ forest estimator for $rms$ bulk velocity}

\subsection{Density-velocity relation in DWT representation}

By definition, a 1-D radial density perturbation $\delta(r)$ is given by
the projection of the 3-D density perturbation $\delta(x,y,r)$ into 1-D,
i.e.
\begin{equation}
\delta(r) = \frac{1}{L_xL_y}\int_{L_x}\int_{L_y} \delta(x,y,r)dxdy=
\frac{1}{L_xL_y\bar{\rho}}
\Bigl{[} \int_{L_x}\int_{L_y} \rho(x,y,r)dxdy - \bar{\rho} \Bigr{]},
\end{equation}
where $L_x \times L_y$ is the area of the coverage of samples on the 
celestial sphere. Both $L_x$ and $L_y$ are larger than the scale 
considered in the radial (or redshift) direction. Since velocity 
field ${\bf v}({\bf x})$ is statistically homogeneous, we have 
$(1/L_xL_y)\int_{L_y} \int_{L_x} (\partial v_x(x,y,r)/\partial x) dx dy
= (1/L_y)\int_{L_y}[ v_x(x_2,y,r)-v_x(x_1,y,r)]/L_x dy \simeq 0$. Similarly,
$(1/L_xL_y)\int_{L_x} \int_{L_y}(\partial v_y/\partial y) dy dx 
\simeq 0$. Thus, subjecting Eq.(1) to the operation
$(1/L_xL_y)\int_{L_x}\int_{L_y}...dxdy$,
we have a 1-D equation
\begin{equation}
\delta(r) = - \frac{1}{H_0f} \frac{d}{dr} v(r),
\end{equation}
where $v(r)$ is a 1-D velocity defined by
\begin{equation}
v(r) =\frac{1}{L_xL_y}\int_{L_x}\int_{L_y} v_r(x,y,r)dxdy
\end{equation}

In redshift space, Eq.(11) is
\begin{equation}
\delta(z)= -\frac{1+z}{H f D}\frac{d}{dz}v(z),
\end{equation}
where $(1+z)$ comes from the expansion of the universe, $v$ is the radial
component of the velocity field, and $D = dr/dz$. For an LCDM model
with $\Omega+\Lambda=1$, we have $D = c [H_0 E(z)]^{-1}$,
$H = H_0 E(z)$, $E(z) =\lbrack\Omega(1+z)^3
+ \Lambda\rbrack^{1/2}$ (Peebles 1993), and
$f \equiv f(z)\approx \lbrack \Omega(1+z)^3 E(z)^{-2}\rbrack^{0.6}$
(Lahav et al. 1991).

The operator $d/dz$ is almost diagonal in wavelet representation
(Farge et al. 1996), and therefore, subjecting Eq.(13) to a DWT with
scaling function $\phi_{jl}(z)$, we have
\begin{equation}
\epsilon^{m}_{jl} \simeq -\frac{1+z}{H f D}
  \sum_{l'=0}^{2^j-1}\epsilon^{v}_{ jl'}
  \int \phi_{jl}(z)\frac{d}{d z}\phi_{jl'}(z)dz
\end{equation}
where $\epsilon^{v}_{ jl}$ is the SFC of the velocity field, given by
$\epsilon^{v}_{jl}=\int v(z) \phi_{jl}(z)d z$. The SFC $\epsilon^{v}_{ jl}$
is proportional to  the bulk velocity $v_j$ on the spatial range given by
the window function $\phi_{jl}$, i.e.
\begin{equation}
v_j=\frac{\int v(z) \phi_{jl}(z)d z}{\int \phi_{jl}(z)d z}=
   \sqrt{\frac{2^j}{L}} \epsilon^{v}_{jl}.
\end{equation}
Since $\int \phi_{jl}(z)\frac{d}{dz}\phi_{jl'}(z)dz=\frac{2^j}{L}
 \Omega^{0,1}_{l-l'}$, where $\Omega^{0,1}_{l-l'}$
is the so called connection coefficient of the basic scaling function
(e.g. Restrepo \& Leaf 1995), Eq.(14) can be rewritten as
\begin{equation}
\epsilon^{m}_{jl}\simeq -\frac{1+z}{H f D}
  \sum_{l'=0}^{2^{j}-1} \frac{2^j}{L}\Omega^{0,1}_{l-l'}
  \epsilon^{v}_{jl'}.
\end{equation}
This is the density-velocity linear Eq.(1) in the DWT 
representation. It is convenient for estimating $\epsilon^{v}_{jl'}$
from $\epsilon^{m}_{jl}$, and {\it vice--versa}. For D4 wavelet
$\Omega^{0,1}_{l-l'}$ is non-zero only for $|l-l'|\leq 2$.

For a Gaussian velocity field, we have
$\langle \epsilon^{v_i}_{j {\bf l'}}
   \epsilon^{v_i}_{ j {\bf l''}}\rangle=0$ if ${\bf l'} \neq {\bf l''}$
(Pando, Feng \& Fang 2001), and therefore, Eq.(16) yields
\begin{equation}
\langle (\epsilon^{m}_{jl})^2\rangle  \simeq
\frac{(1+z)^2}{(H f D)^2}
  \sum_{l'=0}^{2^{j}-1} \left [\frac{2^{j}}{L}\Omega^{0,1}_{l-l'}\right]^2
   \langle (\epsilon^{v}_{jl'})^2 \rangle .
\end{equation}
Moreover, for a uniform and isotropic random field,
$\langle (\epsilon^{m}_{jl})^2\rangle$ and
$\langle (\epsilon^{v}_{jl})^2 \rangle$ are independent
of position index $l$. Eq.(17) gives
\begin{equation}
 \langle (\epsilon^{m}_{jl})^2 \rangle
  \simeq \left [\frac{2^{j}(1+z)}{L H f D}\right]^2
  W \langle (\epsilon^{v}_{jl})^2 \rangle,
\end{equation}
where $W = \sum_{|l-l'|\leq 2}(\Omega^{0,1}_{l-l'})^2$.

\subsection{Results of $rms$ bulk velocity}

The ensemble of $N$ Ly$\alpha$ forests can be seen as a sampling of the
sky with pencil beams. Because the operation
$(1/L_xL_y)\int_{L_x}\int_{L_y}...dxdy$
actually is an average over the celestial sphere, it can be
statistically approximated by an ensemble average over the sky
sampling. That is, the 1-D density perturbation $\delta(r)$ can
statistically be estimated by
\begin{equation}
\delta(r)=\frac{1}{L_xL_y}\int_{L_x}\int_{L_y} \delta(x,y,r)dxdy
  \simeq \frac{1}{N}\sum_{i=1}^{N} \delta(x_i, y_i, r).
\end{equation}
Thus, using Eqs.(7), (8), (9) and (12), we have
\begin{equation}
\epsilon^{m}_{jl}=\int \delta(z)\phi_{jl}(z)dz=
\frac{1}{N}\sum_{i=1}^{N}\int \delta(x_i, y_i, z)\phi_{jl}(z)dz
=-\frac{1}{aA(z)}\epsilon^F_{jl}.
\end{equation}
Substituting Eqs.(20) and (15) into Eq.(18), we finally have the
estimator for the $rms$ bulk velocity on scale $j$,
$\sigma^{v_r}_j=\langle v^2_j \rangle^{1/2}$, of the underlying
mass field of Ly$\alpha$ forests
\begin{equation}
\sigma^{v_r}_j \simeq \frac{R f(z)}{aA(z)W^{1/2}(1+z)}\lbrack 2^{J-j}
  \langle (\epsilon^{F}_{jl})^2 \rangle \rbrack^{1/2},
\end{equation}
where $R = cL/2^J$ is the pixel size in km s$^{-1}$, and the physical
length scale corresponding to $j$ is $cL[2^jH_0E(z)]^{-1}$. The 
superscript $v_r$ is to remind that it is a projection along the 
$r$-direction. The $rms$
bulk velocity on a given scale is then determined from the flux
fluctuations $\delta F$ on the same scale.

Using the estimator Eq.(21), we calculate the $rms$ bulk velocities
for the three redshift regions $z\pm 0.25$ with $z$= 2.25, 2.75, and
3.25. The results are presented in Table 1. The scales in Table 1 have
the same meaning as those in Fig. 1. In this calculation $a$ is taken to be
$1.6$, and $R=293$ km s$^{-1}$. The factor $A(z)$ is calculated from
the mean transmission $\langle F(z) \rangle$ in each redshift range
via $A(z) = -\ln \langle F(z) \rangle $. 

It should be pointed out that the average over $N$ pencil beams in eq.(19) will cause 
a Poisson error, which then contributes to the variance of $\delta(r)$. This problem
is the same as the Poisson correction in calculating the power spectrum 
of galaxies (Peebles 1980, Fang \& Feng 2000). In Table 1 this correction 
has already been applied.

\begin{table}
\caption{1-D $rms$ bulk velocities $\sigma^{v_r}_j$ (km s$^{-1}$)
\label{table-1}}
\bigskip
\begin{tabular}{cccc}
\tableline
Scales (h$^{-1}$Mpc) & 92 & 46 &  23  \\
\tableline
  z = 2.25    & 130$\pm$50 & 190$\pm$55 & 230$\pm$50  \\
  z = 2.75    & 62$\pm$40  & 86$\pm$30  & 94$\pm$20  \\
  z = 3.25    & 110$\pm$45 & 170$\pm$50 & 190$\pm$40 \\
\tableline
\end{tabular}
\end{table}

The errors in Table 1 are calculated from different realizations of 
phase-randomized spectra which preserves the Gaussianity on large 
scales (Jamkhedkar, Zhan, \& Fang 2000; Zhan, Jamkhedkar, \& Fang 2001).
It is essentially similar to the bootstrap error estimation
(Mo, Jing, \& B\"orner 1992). Besides this error, there are also systematic
errors caused by the uncertainties of the factor $aA(z)$ in Eq.(7). For
a given redshift range, the uncertainty of the mean transmission $A(z)$ is
less than 10\%, and the uncertainty of $a$ is about (1.9-1.5)/1.6 = 25\%.
Therefore, the total systematic error would be about 30\%.

Continuum--fitting may cause error if the continua $F_c$ of the 60
QSOs fluctuate significantly on the same scale considered, say 92
h$^{-1}$ Mpc. However, the PDFs of $\epsilon^F_{jl}$ on large scales
are perfectly Gaussian. Therefore, the SFC $\epsilon^F_{jl}$ may be
significantly contaminated by the continuum fluctuations {\it only if}
the PDF of the continuum fluctuations is also Gaussian. This seems
quite unlikely, as the emission from each QSO is highly non-thermal.

All the 1-D $rms$ bulk velocities listed in Table 1 are consistent
with the assumption in \S 2, i.e. the peculiar velocity $v_r$ is
much smaller than the spatial size considered, i.e.
$H_0r \geq 2300$ km s$^{-1}$. Moreover, the error from the resolution
$R$ of the samples is also small, as $R$ is also much less than the 
spatial size
considered. In other words, the position and density uncertainties
caused by $v_r$ in Eq.(2) and $R$ are negligible in comparison with the
scales listed in Table 1. We should point out that the result of
Table 1 is not sensitive to the choice of wavelets. Our analysis relies
only on the orthogonality, completeness, and locality of the scaling
function. Therefore, all wavelets with compactly supported basis will
produce similar results.

As expected, Table 1 shows that $\sigma_v$ decreases with
scales from 23 to 92 h$^{-1}$Mpc, and it also decreases with redshift from
2.25 to 3.25. However, there is a dip of the bulk velocities on
all scales at around z = 2.75. This should be verified with
other sets of Ly$\alpha$ QSO transmission fluxes.

\section{Discussions and conclusions}

To properly compare the estimated $\sigma^{v_r}_j$ with other observed or 
theoretical results, we should emphasize that $\sigma^{v_r}_j$ and $v_j$ are
defined by the DWT decomposition of velocity field (Eq.[15]), not by the
velocities of particles or galaxies. It has been point out recently that
the velocity defined by the particle-counting or galaxy-counting method may
have different statistical properties from that defined by velocity
field decomposition (Yang et al. 2001.) Therefore, we should compare
our results with those that are also based on field decomposition if
available.

 From Eqs.(12) and (15), we have
\begin{equation}
v_j=\frac{\int_{L_x}\int_{L_y}\int v_r(x,y,z) \phi_{jl}(z)dxdyd z}
     {L_xL_y\int \phi_{jl}(z)d z}.
\end{equation}
In terms of the DWT analysis, the average $(1/L_x)\int_{L_x} ... dx$
can be replaced by scaling function projection window
$\int ... \phi_{j_x,l}(x)dx/\int\phi_{j_x,l}(x)dx $, where
$j_x$ corresponds to scale $L_x$.  Thus, Eq.(22) can be rewritten as
\begin{equation}
v_j = \frac{\int v_r(x,y,z)\phi_{j_x,l_x}(x)\phi_{j_y,l_y}(y)
\phi_{jl}(z)dxdydz}
     {\int \phi_{j_x,l_x}(x)\phi_{j_y,l_y}(y)
\phi_{jl}(z)dxdydz}=v_{j_x,j_y,j}.
\end{equation}
Since the scales $L_x$ and $L_y$ are much larger than the radial scale 
$L/2^j$ considered, we have $j_x$,$j_y$ $\ll j$. Therefore, 
$\sigma^{v_r}_j$ actually is the 1-D $rms$ bulk velocity of 
the 3-D mode $(j_x,j_y,j)$ with $j_x$,$j_y$ $\ll j$, i.e.  
$\sigma^{v_r}_j=\langle v^2_{j_x,j_y,j}\rangle^{1/2}
=\sigma^{v_r}_{j_x,j_y,j}$.
For a homogeneous and isotropic random mass field with a relatively flat
power spectrum around 100-150 Mpc, we have
$\sigma^{v_r}_{j,j,j}\simeq 
\sqrt{3}\sigma^{v_r}_{j_x,j_y,j}$ if $j_x,j_y \ll j$,
or equivalently, $L_x$ and $L_y$ are much larger than the physical scales
of $j$ (Yang et al 2001).

Since the results listed in Table 1 are derived for high redshifts, there
is no directly measured result available for comparison. The
bulk velocity in a sphere of radius 50 h$^{-1}$Mpc around the Local Group
is found to be $370\pm 110$ km s$^{-1}$ (Dekel et al 1998). The scale of
the sphere is comparable to the linear scale 92 h$^{-1}$Mpc in Table 1.
If this result can be considered as the bulk velocity of a box
100$^3$ h$^{-3}$Mpc$^{3}$, the corresponding 1-D bulk velocity $v_j$
on scale 100 h$^{-1}$Mpc (with $L_x, L_y \gg 100$h$^{-1}$Mpc) 
is about $370/\sqrt{3}/\sqrt{3}=123$ km s$^{-1}$ (the first $\sqrt{3}$ is due
to the transform from 3-D to 1-D of the velocity, and the second $\sqrt{3}$ 
is due to the transform from mode $(j,j,j)$ to $(0,0,j)$). Considering 
linear redshift evolution, this 1-D bulk velocity yields 99, 92 and 88 
km s$^{-1}$ at redshift $z=$2.25, 2.75, and 3.25, respectively. These 
results are consistent with that shown in Table 1. Nevertheless we should 
remember that the bulk velocity 370 km s$^{-1}$ is from one realization, 
not $rms$ value, and it may have the uncertainties of galaxy-counting as 
well as galaxy bias.

The 1-D $rms$ bulk velocity in the DWT modes $(j,j,j)$ at redshift $z=0$
has been calculated from N body simulations. The results on 92 h$^{-1}$Mpc
are 60$\pm$10 km s$^{-1}$, 70$\pm$20 km s$^{-1}$, and 80$\pm$20 km s$^{-1}$ 
for the SCDM, LCDM and $\tau$CDM models respectively (Yang et al 2001). 
All models give lower $rms$ bulk velocity than that shown in Table 1.
One cannot, however, conclude that the N body simulated result is 
inconsistent with that shown in Table 1. This is because the $rms$ bulk 
velocity is sensitive to the perturbation on large scales, while the 
simulation box in Yang et al (2001) is 256$^3$ h$^{-3}$Mpc$^3$. It has 
been pointed out (Tormen \& Bertschinger 1996) that the linear 3-D $rms$ 
bulk velocity of a cube of side 100 Mpc is well over 500 km s$^{-1}$ for 
an SCDM model with $\sigma_8=1$. Correspondingly, for an LCDM model with
$\sigma_8=0.7$ the 1-D $rms$ bulk velocities of DWT mode 
$(L_x,L_y, 92$ h$^{-1}$Mpc), i.e. $\sigma^{v_r}_{0,0,j}$, are 
110 km s$^{-1}$, 105 km s$^{-1}$, and 100 km s$^{-1}$ at $z=$2.25, 2.75, 
and 3.25 respectively. Therefore, the 
results shown in Table 1 basically are consistent with the linear 
evolution theory within 1$\sigma$ confidence level.

We should also emphasize that the average over a
QSO ensemble, i.e. average over the celestial sphere, is important.
The errors in Table 1 also comes from the fact that the number of
Ly$\alpha$ forests in a given redshift bin is still small (large Poisson 
correction caused by eq.[19]). Nevertheless,
one can already conclude that a set of Ly$\alpha$ forests can provide a
valuable estimate of the $rms$ bulk velocity of the underlying mass
field on scales from a few tens to $\lesssim$ 100 h$^{-1}$Mpc. With a
large set of Ly$\alpha$ forests, one may map the bulk velocity on large
scales in a wide range of redshift, and then, set a robust constraint on
the power spectrum on large scales.

\acknowledgments
The authors would like to thank David Burstein for helpful suggestions.

\newpage

\plotone{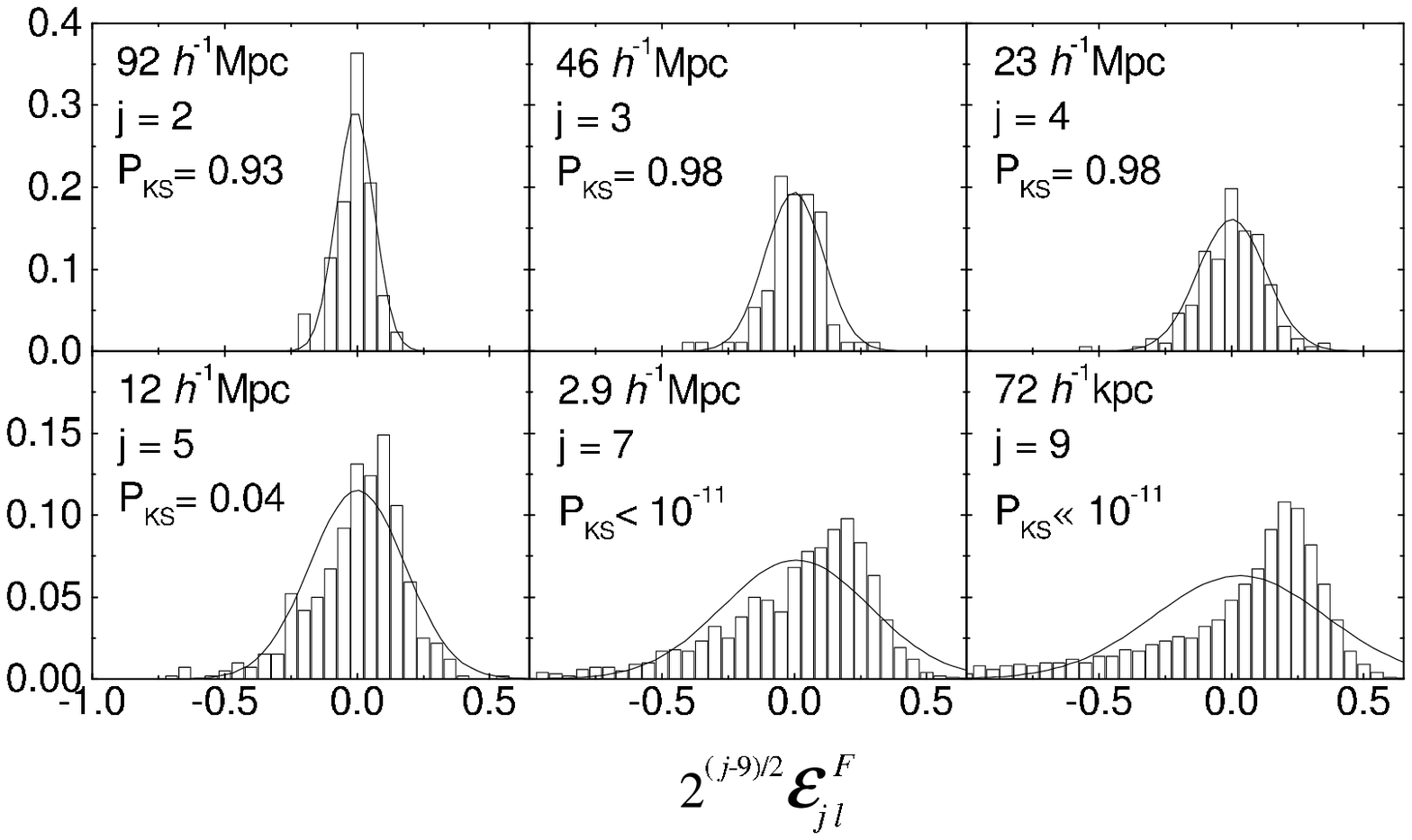}
\figcaption {Probability distribution functions of $\epsilon^{F}_{jl}$s of
the data set consisting of all QSO spectrum segments within the
redshift range from 2.5 to 3.0.  An LCDM ($\Omega = 0.3$ and $\Lambda=0.7$)
universe is assumed to calculate the comoving length scales. P$_{KS}$ is
the K-S probability of the PDF being Gaussian. } \label{Fig1}

\end{document}